\documentclass[%
reprint,
superscriptaddress,
amsmath,amssymb,
prb,
longbibliography
]{revtex4-2}

\usepackage{color}
\usepackage{xcolor}
\usepackage{graphicx}
\usepackage{dcolumn}
\usepackage{siunitx}
\usepackage{bm}
\usepackage{hyperref}
\hypersetup{colorlinks=true,linkcolor=blue,citecolor=blue,urlcolor=blue}
\usepackage{soul} 
\usepackage[colorinlistoftodos]{todonotes}
\usepackage{mhchem}

\begin{document}
\title{Multiband-Induced Complex Magnetotransport in the Kagome Metal LuNb$_6$Sn$_6$}

\author{Hongye Zhang}
\thanks{These authors contributed equally to this work. }
\affiliation{Low Temperature Physics Laboratory, School of Physics \& Center of Quantum
Materials and Devices, Chongqing University, Chongqing 401331, China}

\author{Xingyu Ma}
\thanks{These authors contributed equally to this work. }
\affiliation{Beijing National Laboratory for Condensed Matter Physics and Institute of Physics,
Chinese Academy of Sciences, Beijing 100190, China}
\affiliation{Department of Physics and Institute of Theoretical Physics, University of Science and Technology Beijing, Beijing 100083, China}

\author{Ying Zhu}
\thanks{These authors contributed equally to this work. }
\affiliation{Low Temperature Physics Laboratory, School of Physics \& Center of Quantum Materials and Devices, Chongqing University, Chongqing 401331, China}

\author{Jingyang Zhu}
\thanks{These authors contributed equally to this work. }
\affiliation{Low Temperature Physics Laboratory, School of Physics \& Center of Quantum Materials and Devices, Chongqing University, Chongqing 401331, China}

\author{Andreas Eich}
\affiliation{Institute for Quantum Materials and Technologies, Karlsruhe Institute of Technology, Kaiserstraße 12, 76131 Karlsruhe, Germany}

\author{Yuan Tang}
\affiliation{Low Temperature Physics Laboratory, School of Physics \& Center of Quantum Materials and Devices, Chongqing University, Chongqing 401331, China}

\author{Xiong Wang}
\affiliation{Low Temperature Physics Laboratory, School of Physics \& Center of Quantum Materials and Devices, Chongqing University, Chongqing 401331, China}

\author{Ningxi Yang}
\affiliation{Low Temperature Physics Laboratory, School of Physics \& Center of Quantum Materials and Devices, Chongqing University, Chongqing 401331, China}

\author{Aifeng Wang}
\affiliation{Low Temperature Physics Laboratory, School of Physics \& Center of Quantum
Materials and Devices, Chongqing University, Chongqing 401331, China}

\author{Yisheng Chai}
\affiliation{Low Temperature Physics Laboratory, School of Physics \& Center of Quantum
Materials and Devices, Chongqing University, Chongqing 401331, China}

\author{Xinrun Mi}
\email{xinrunmi@cqu.edu.cn}
\affiliation{Low Temperature Physics Laboratory, School of Physics \& Center of Quantum Materials and Devices, Chongqing University, Chongqing 401331, China}
\affiliation{Chongqing Police College, Chongqing 401331, China}

\author{Michael Merz}
\email{michael.merz@kit.edu}
\affiliation{Institute for Quantum Materials and Technologies, Karlsruhe Institute of Technology, Kaiserstraße 12, 76131 Karlsruhe, Germany}
\affiliation{Karlsruhe Nano Micro Facility, Karlsruhe Institute of Technology, Kaiserstraße 12, 76131 Karlsruhe, Germany}

\author{Quansheng Wu}
\email{quansheng.wu@iphy.ac.cn}
\affiliation{Beijing National Laboratory for Condensed Matter Physics and Institute of Physics,
Chinese Academy of Sciences, Beijing 100190, China}
\affiliation{University of Chinese Academy of Sciences, Beijing 100049, China}
 
\author{Mingquan He}
\email{mingquan.he@cqu.edu.cn}
\affiliation{Low Temperature Physics Laboratory, School of Physics \& Center of Quantum
Materials and Devices, Chongqing University, Chongqing 401331, China}
\date{\today}

\begin{abstract}

We report the electrical transport properties of the kagome metal LuNb$_6$Sn$_6$, which undergoes a charge density wave (CDW) transition at $T_\mathrm{CDW}\approx80$ K. Similar to the kagome metals \textit{A}\ce{V3Sb5} (\textit{A} = K, Rb, Cs), LuNb$_6$Sn$_6$ develops complex magnetotransport behavior near and below the CDW transition. In particular, the longitudinal magnetoconductivity exhibits an approximately linear field dependence below about 1 T, whereas the field dependence of Hall conductivity displays pronounced nonlinearity and a field-induced sign reversal. Although an anomalous Hall-like contribution can be extracted by subtracting a local background from the Hall conductivity, mobility-spectrum analysis shows that the overall magnetotransport can be largely accounted for by multiband conduction involving high-mobility electronlike and holelike carriers. Nevertheless, sizable deviations from the multiband model emerge near $T_{\mathrm{CDW}}$, pointing to additional CDW-related scattering processes. These results demonstrate that high-mobility multiband transport, together with CDW-induced modifications of carrier scattering, play important roles in the complex magnetotransport of LuNb$_6$Sn$_6$.
\end{abstract}

\maketitle

\section{Introduction}

\begin{figure*}
\centering
\includegraphics[scale=0.55]{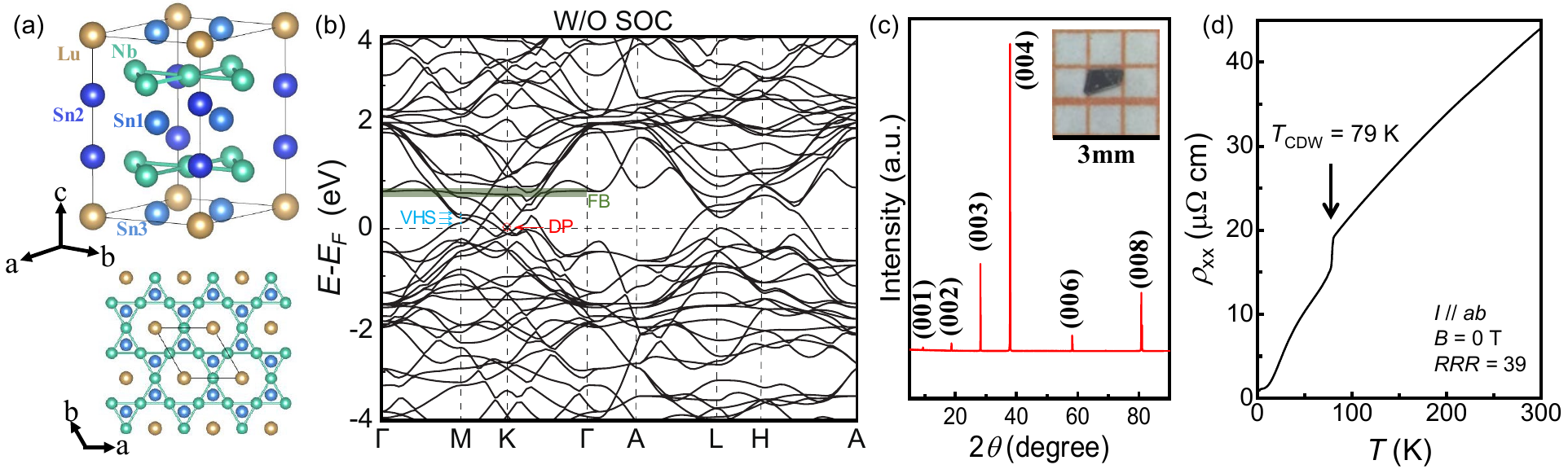}
\caption{(a) Upper panel: Side view of the crystal structure of \ce{LuNb6Sn6} (space group: $P6/mmm$), showing the Nb-Nb bonds. The different Sn sites are labeled as Sn1, Sn2, and Sn3. Lower panel: Top view of the crystal structure along the $c$ axis. The Nb atoms form a kagome lattice. (b) Calculated electronic band structure of \ce{LuNb6Sn6} in the high-temperature pristine phase without spin-orbit coupling (SOC). Typical features including a flat band (FB), van Hove singularities (VHS) and a Dirac point (DP) near the Fermi level are highlighted out.  (c) X-ray diffraction pattern of a typical \ce{LuNb6Sn6} single crystal, with clear identification of the ($00l$) peaks. The inset shows a photograph of a typical \ce{LuNb6Sn6} single crystal. (d) Temperature dependence of the longitudinal resistivity.}
\label{fig1}
\end{figure*}

The kagome lattice, a two-dimensional network of corner-sharing triangles, has been extensively studied over the past decades \cite{Syozi1951}. The geometric frustration inherent to this lattice gives rise to a characteristic electronic structure with flat bands, saddle points associated with van Hove singularities, and Dirac cones \cite{Neupert2022}. The discovery of superconductivity and unusual charge-density-wave (CDW) order in the kagome metals \textit{A}\ce{V3Sb5} (\textit{A} = K, Rb, Cs) has provided a fertile platform for studying the interplay among lattice geometry, band topology, and electronic correlations \cite{Ortiz2019, Ortiz2020, Ortiz2021, Yin2021}. In particular, the CDW phase has been proposed to be associated with chiral charge order and intrinsic time-reversal symmetry breaking, although its microscopic origin remains under debate \cite{Jiang2021a, Wang2021CDW, Shumiya2021, Yang2020, Yu2021, Zheng2021_gating, Chendong2021, Zhou2021, Mielke2021, Yu2021b, Gan2021, Mi_2022, Liang2021CDW, Zhao2021a, Chen2021rotonpair, Xu2021, Li2021a, Hu2022co}.

One important indication of possible time-reversal symmetry breaking in the CDW phase of \textit{A}\ce{V3Sb5} is the observation of an anomalous Hall effect (AHE)-like Hall conductivity \cite{Yang2020, Yu2021, Zheng2021_gating, Zhou2021}. However, extrinsic mechanisms, such as skew scattering from spin clusters or ferromagnetic fluctuations, may also contribute to such an AHE-like response \cite{Yang2020, Yu2021, Zheng2021_gating}. Moreover, the AHE-like feature is usually obtained indirectly by subtracting a local background from the raw Hall conductivity, which itself shows complex field dependence. Specifically, the Hall conductivity exhibits a field-induced sign reversal, whereas the longitudinal magnetoconductivity varies nearly linearly with magnetic field below about 1 T \cite{Koshelev}. Given the multiband electronic structure of these kagome metals, such complex magnetotransport can in principle arise from the underlying Fermiology without invoking additional nontrivial mechanisms \cite{Zhang_complex}. Indeed, the magnetotransport properties of \textit{A}\ce{V3Sb5} can be largely accounted for by two-band analysis \cite{Gan2021, Mi_2022, Mi2023}, and similar behavior has also been found in the kagome systems \textit{A}\ce{Ti3Bi5} \cite{Chen_Ti3Bi5} and \ce{EuTi3Bi4} \cite{Shu_2025} in the absence of charge order. More recent work suggested that the complex magnetotransport of \textit{A}\ce{V3Sb5} can be described semi-quantitatively by concave Fermi pockets near the van Hove singularities, emphasizing the importance of band structure in determining transport properties \cite{Koshelev}. Mobility-spectrum analysis further revealed that multiband transport, particularly tiny Fermi pockets with high carrier mobility, is crucial for the complex magnetotransport behavior in \textit{A}\ce{V3Sb5} \cite{Liu_MS}.   

Beyond \textit{A}\ce{V3Sb5}, CDW order has also been found in the kagome metal \ce{ScV6Sn6} below $T_\mathrm{CDW}\sim90$ K \cite{Arachchige_ScVSn}. Unlike \textit{A}\ce{V3Sb5}, where the CDW is associated with in-plane lattice distortions of the V-kagome layer, the CDW in \ce{ScV6Sn6} is driven by out-of-plane distortions along the Sc-Sn chain \cite{Arachchige_ScVSn, Meier_rattle,Hu2024,Pok_scVSn,Cao2023,Lee2024,Korshunov2023}. Despite these differences, signatures of hidden magnetism and AHE-like transport behavior have also been found in \ce{ScV6Sn6} \cite{Guguchia2023_ScVsn,Yi_ScVSn_AHE,Mo_Sc}. These findings suggest that charge order associated with time-reversal symmetry breaking, possibly related to orbital loop currents, may be a generic feature of these kagome systems \cite{Guguchia2023_ScVsn,Grandi_AHE,Julian_ScVSn}. On the other hand, simple two-band modeling can describe the complex Hall conductivity of \ce{ScV6Sn6}, including the field-induced sign reversal \cite{DeStefano_Sc}. Compared with \textit{A}\ce{V3Sb5}, the role of the multiband electronic structure in the complex magnetotransport of \ce{ScV6Sn6} remains less explored. Interestingly, recent studies have shown that the kagome metal \ce{LuNb6Sn6}, an isostructural compound of \ce{ScV6Sn6}, also develops a similar CDW order below $T_\mathrm{CDW}\sim70$ K, characterized by out-of-plane displacements of the Lu-Sn chain \cite{Ortiz_LuNb6Sn6,Rui_LuNbSn,Yang_LuNbSn,Meier_LuNbSn,Philp_LuNbSn}. This material therefore provides an ideal platform for examining unusual magnetotransport in kagome metals with charge order. Whether \ce{LuNb6Sn6} exhibits a similar AHE-like anomaly, and how multiband transport contributes to its magnetotransport properties, remain open questions.

\begin{table}[b]
    \centering
    \caption{Crystallographic data for \ce{LuNb6Sn6} derived from single-crystal XRD (SC-XRD) measurements at 300 K.\@ Errors shown are statistical errors from the refinement.
   }
    \label{table1}
    \begin{tabular}{ll}
        \hline\hline
        Parameters              & SCXRD\\
        \hline
        Chemical formula        & \ce{LuNb6Sn6} \\
        Formula weight (g/mol)  & 1444.67 \\
        X-ray density (g/cm³)   & 8.8047  \\
        Space group             & $P6/mmm$ (No.\,191) \\
        $Z$                     & 1 \\
        $a$ (\AA)               & 5.7506(1) \\
        $b$ (\AA)               & 5.7506(1) \\
        $c$ (\AA)               & 9.5125(1) \\
        $\alpha$                & \ang{90} \\
        $\beta$                 & \ang{90} \\
        $\gamma$                & \ang{120} \\
        $V$ (\AA$^3$)           & 272.4 \\ \hline
        $R_1$ (\%)          & 1.87 \\
        $wR_2$ (\%)             & 5.22\\
        GOF                & 1.87 \\
        \hline
    \end{tabular}
\end{table}

In this article, we study the magnetotransport properties of single-crystalline \ce{LuNb6Sn6}, which enters the CDW state below $T_\mathrm{CDW}=79$ K. Complex magnetotransport develops below $\sim100$ K, and is characterized by a nearly linear field-dependent longitudinal magnetoconductivity below $\sim1$ T and a field-induced sign reversal in the Hall conductivity. Similar to \textit{A}\ce{V3Sb5} and \ce{ScV6Sn6}, an AHE-like feature can be extracted after subtracting a local background from the Hall conductivity. Using mobility-spectrum analysis, we show that both the complex longitudinal magnetoconductivity and Hall conductivity can be largely captured by Drude-type multiband transport. These results highlight the central role of multiband conduction, especially Fermi pockets with high mobility, in governing complex magnetotransport in kagome metals.

\section{Method}
Single crystals of \ce{LuNb6Sn6} were grown by the Sn self-flux method. High-purity Lu, Nb, and Sn with purities of 99.9\%, 99.99\%, and 99.999\%, respectively, were mixed in a molar ratio of Lu:Nb:Sn = 8:2:90. The mixture was placed in an alumina crucible and sealed under high vacuum in a quartz tube. The sealed ampoule was heated to 1150 $^{\circ}$C, kept at this temperature for 30 hours, and then slowly cooled to 900 $^{\circ}$C at a rate of 1 $^{\circ}$C/hour. Millimeter-scale single crystals of \ce{LuNb6Sn6} [see the inset of Fig.~\ref{fig1}(c)] were obtained after removing the excess Sn flux using a centrifuge \cite{Ortiz_LuNb6Sn6}.  

The crystal structure and precise determination of the composition was probed by single-crystal  x-ray diffraction (SC-XRD). SC-XRD measurements were performed at 300 K on a high-flux, high-resolution, rotating anode Rigaku Synergy-DW (Mo/Ag) diffractometer using Mo K$_{\alpha}$ radiation ($\lambda$ = 0.7107 \AA).\@ The measured specimens had a size of $\approx$ 80 $\times$ 70 $\times$ 60  µm$^3$.\@ Data collection, integration of the intensities, and data reduction were performed using the most recent version of the \textsc{CrysAlisPro} software package~\cite{Crysalis}.\@ The crystal structure was solved using the charge-flipping algorithm implemented in \textsc{Superflip}~\cite{palatinus2007superflip} and subsequently refined with \textsc{JANA2006}~\cite{Jana2006}.

Longitudinal resistivity and Hall resistivity measurements were performed using a conventional Hall-bar geometry in a Physical Property Measurement System (PPMS, Quantum Design Dynacool 9 T). To minimize the effects of contact misalignment, the longitudinal magnetoresistivity $\rho_{xx}(B)=[\rho_{xx}(+B)+\rho_{xx}(-B)]/2$ was symmetrized and the Hall resistivity $\rho_{yx}(B)=[\rho_{yx}(+B)-\rho_{yx}(-B)]/2$ was antisymmetrized with respect to magnetic field.

First-principles calculations were performed within density functional theory (DFT) using the Vienna \textit{ab initio} Simulation Package (VASP) \cite{Kresse1996}. The projector augmented-wave (PAW) method was employed to describe the electron--ion interactions \cite{Kresse1999}, and the exchange--correlation functional was treated within the generalized gradient approximation (GGA) using the Perdew--Burke--Ernzerhof (PBE) parametrization \cite{Perdew1996}. The plane-wave kinetic-energy cutoff was set to 440~eV, and the Brillouin zone was sampled using a $5\times5\times6$ $\Gamma$-centered $k$-point mesh. The electronic self-consistency criterion was set to $10^{-8}$~eV, with a Gaussian smearing width of 0.05~eV. The calculations were performed without spin polarization or spin--orbit coupling.

\begin{figure*}
\centering  
\includegraphics[scale=0.6]{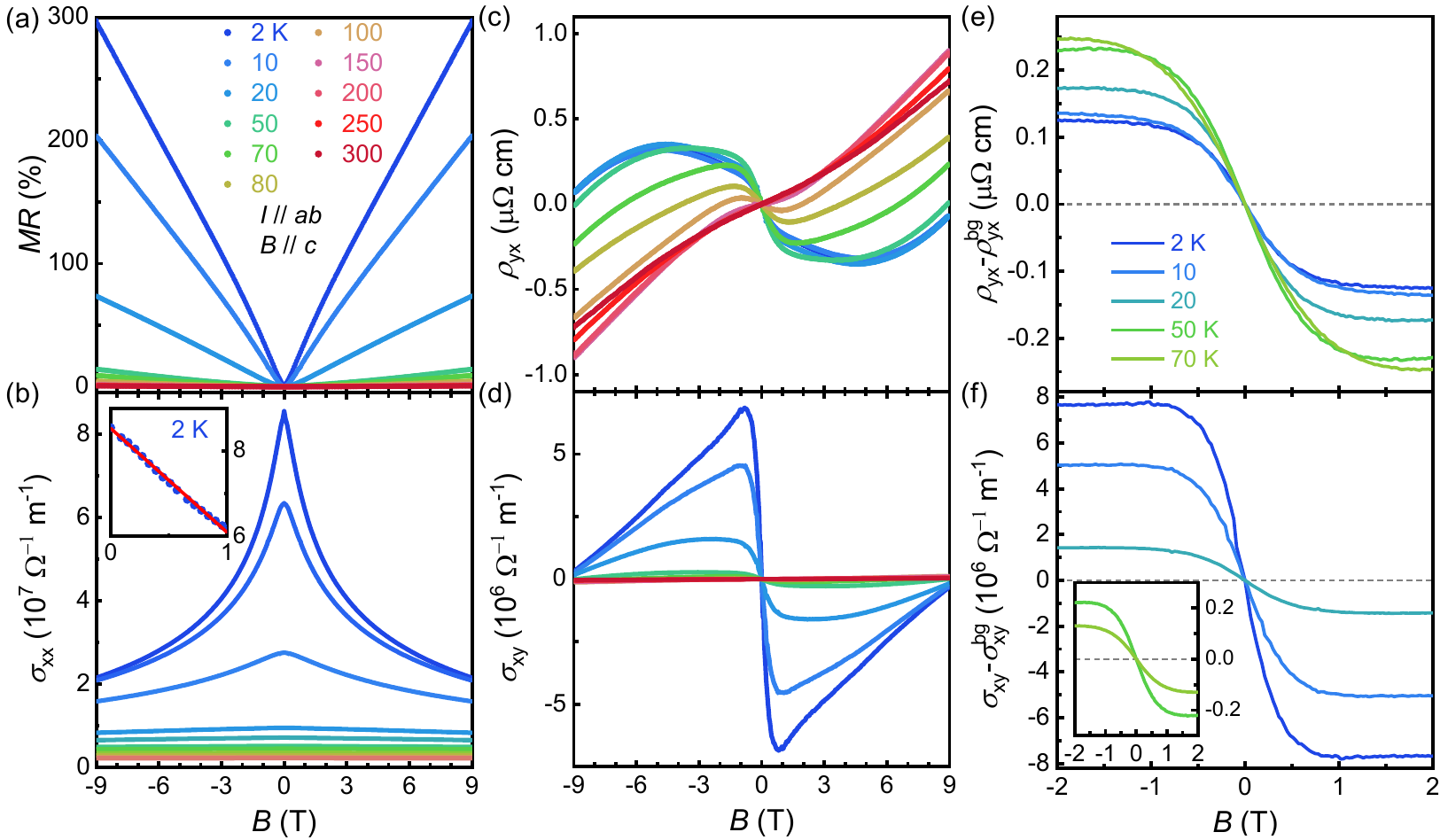}
\caption{(a) Magnetoresistance (MR) and (b) magnetoconductivity of \ce{LuNb6Sn6} measured at selected temperatures. (c) Hall resistivity and (d) Hall conductivity of \ce{LuNb6Sn6} recorded at the corresponding temperatures shown in (a) and (b). The inset in (b) shows a linear fitting (red solid line) to the low field magnetoconductivity measured at 2 K. (e) and (f) Low-temperature Hall resistivity and Hall conductivity obtained after subtracting the local linear background ($\rho_{yx}^{bg}$, $\sigma_{xy}^{bg}$).}
\label{fig2}
\end{figure*}

\section{Results and Discussion}

\begin{table}
    \centering
    \caption{ Atomic coordinates, Wyckoff positions, positional occupation, and equivalent displacement parameters, $U_{\mathrm{eq}}$,\@ of \ce{LuNb6Sn6}.}
    \label{table2}
    \begin{tabular}{lcccccc}
        \hline \hline
        Atom & Wyck. & \hspace*{2mm} $x$ \hspace*{2mm}  & \hspace*{2mm} $y$ \hspace*{2mm} & $z$ & Occ. & $U_{\mathrm{eq}}$ (\AA$^{2}$) \\
        Lu  & $1a$ & 0     & 0     & 0        & 1     & 0.010948(35) \\
        Sn1  & $2c$ & 2/3   & 1/3   & 0  & 1 & 0.005341(21) \\
        Sn2  & $2d$ & 2/3   & 1/3   & 1/2  & 1 & 0.006226(23) \\
        Sn3  & $2e$ & 0    & 0     & 0.327212(27)  & 1     & 0.009816(37) \\
        Nb  & $6i$ & 1/2    & 1/2     & 0.248865(12)  & 1     & 0.003884(25) \\ \hline \hline
    \end{tabular}
\end{table}

Figure~\ref{fig1}(a) shows the crystal structure of high-temperature pristine \ce{LuNb6Sn6} in the absence of CDW order. Similar to \ce{ScV6Sn6}, \ce{LuNb6Sn6} crystallizes in the \ce{HfFe6Ge6}-type hexagonal structure (space group No. 191, $P6/mmm$), in which each unit cell contains two kagome layers formed by Nb atoms within the $ab$ plane. Alternating \ce{LuSn2}, kagome \ce{Nb3Sn}, and Sn layers are stacked along the $c$ axis. As shown in Fig.~\ref{fig1}(b), typical kagome-lattice features are revealed in the calculated electronic band structure of \ce{LuNb6Sn6}. In particular, a flat band is identified around 0.78 eV along the $\Gamma-\mathrm{M}-\mathrm{K}-\Gamma$ path. Moreover, a van Hove singularity (VHS) at M and a Dirac-like band at K appear near the Fermi energy $E_\mathrm{F}$, while additional VHS and Dirac-like features located slightly away from $E_\mathrm{F}$ can also be identified. These features have been clearly revealed by angle resolved photoemission spectroscopy (ARPES) experiments \cite{Ortiz_LuNb6Sn6,Rui_LuNbSn}. In addition, the electronic structure displays a clear multiband character, with several bands crossing the Fermi level. A large circular pocket centered at the $\overline{\Gamma}$ point, a large triangular
and a small circular pockets around the $\overline{\mathrm{K}}$ point have been identified by ARPES \cite{Ortiz_LuNb6Sn6,Rui_LuNbSn,Yang_LuNbSn}. Quantum oscillation experiments also revealed two dominant frequencies, $F_\alpha\sim10-20$ T and $F_\beta\sim150-200$ T, together with additional high frequencies up to 10 kT \cite{Philp_LuNbSn,Beek_LuNbSn}. As discussed below, this multiband electronic structure plays an important role in the transport properties of \ce{LuNb6Sn6}.          

Figure~\ref{fig1}(c) presents the single-crystal X-ray diffraction pattern of \ce{LuNb6Sn6}. The ($00l$) peaks are clearly resolved, indicating the [001] preferred orientation of the as-grown crystal.  Our SC-XRD data show that the samples are single-phase and of high quality. The refined crystallographic data is presented in Table~\ref{table1}. During the refinement, the occupancy parameters of the individual Wyckoff positions were also allowed to vary. As shown in Table~\ref{table2}, all positions remain fully occupied despite this freedom, ruling out structural disorder and partial site occupancy.  The longitudinal electrical resistivity $\rho_{xx}$ measured in zero magnetic field, shown in Fig.~\ref{fig1}(d), exhibits overall metallic behavior with a residual resistivity ratio RRR = $\rho_{xx}$(300 K)/$\rho_{xx}$(5 K)=39. In addition, a sharp jump occurs at $T_\mathrm{CDW}=79$ K, signaling the CDW transition. The transition temperature appears to be sensitive to sample quality. Early studies reported a lower transition temperature of $T_\mathrm{CDW}=68$ K \cite{Ortiz_LuNb6Sn6,Yang_LuNbSn,Meier_LuNbSn}, while more recent work using an improved growth method found higher values of $T_\mathrm{CDW}=84$ K \cite{Beek_LuNbSn}. The relatively high CDW transition temperature observed here suggests the high quality of our single crystals.

In Fig.~\ref{fig2}, we present the magnetotransport properties of \ce{LuNb6Sn6} with the electrical current applied within the $ab$ plane and the magnetic field applied along the $c$ axis. Positive longitudinal magnetoresistance, MR$=[\rho_{xx}(B)-\rho_{xx}(0)]/\rho_{xx}(0)$, is observed over the measured temperature range. At high temperatures above about 50 K, $\rho_{xx}(B)$ shows a typical quadratic field dependence. Below 50 K, however, $\rho_{xx}(B)$ evolves toward an almost linear field dependence. No saturation is reached up to 9 T, and a large MR of approximately 200\% is obtained at 2 K and 9 T. As shown in Fig.~\ref{fig2}(b), this peculiar behavior is also seen in the magnetoconductivity $\sigma_{xx}(B)=\rho_{xx}(B)/[\rho_{xx}^2(B)+\rho^2_{yx}(B)]$, where $\rho_{yx}(B)$ is the Hall resistivity [see Fig.~\ref{fig2}(c)]. Below 50 K, $\sigma_{xx}(B)$ shows an approximately linear field dependence below about 1 T [see the inset of Fig.~\ref{fig2}(b)]. Between 2 and 4 T, $\sigma_{xx}(B)$ exhibits an upward curvature, and at higher magnetic fields it recovers a quasi-linear behavior. Similar complex field dependences of the longitudinal magnetoconductivity have also been reported in \textit{A}\ce{V3Sb5} \cite{Koshelev} and \ce{EuTi3Bi4} \cite{Shu_2025}.

The Hall resistivity and Hall conductivity show even richer behavior, as displayed in Figs.~\ref{fig2}(c) and \ref{fig2}(d). Above 150 K, the Hall resistivity $\rho_{yx}(B)$ is dominated by a positive slope with a slight nonlinear field dependence, suggesting that the high-temperature Hall response is mainly governed by holelike carriers with multiband contributions. Upon cooling below 100 K, the nonlinear field dependence of $\rho_{yx}(B)$ becomes more pronounced, developing a ``Z''-shaped profile and even a field-induced sign reversal. At small magnetic fields below about 1 T, $\rho_{yx}(B)$ initially exhibits a negative slope, which changes to a positive slope at higher fields and produces a sign change at a temperature-dependent magnetic field. Both the ``Z''-shaped feature and the sign-change behavior become more pronounced upon cooling. Below 50 K, $\rho_{yx}(B)$ becomes even more complex. Its slope changes twice, around 1 and 6 T, and a positive slope is recovered only above about 6 T. These complex field-dependent features are also manifested in the Hall conductivity $\sigma_{xy}(B)=\rho_{yx}(B)/[\rho_{xx}^2(B)+\rho^2_{yx}(B)]$. By subtracting a local linear background, an AHE-like Hall response can be resolved, similar to that observed in \textit{A}\ce{V3Sb5} and \ce{ScV6Sn6} [see Figs.~\ref{fig2}(e) and \ref{fig2}(f)]. Since long-range magnetic order is absent in \ce{LuNb6Sn6}, as in \textit{A}\ce{V3Sb5} and \ce{ScV6Sn6}, this AHE-like Hall feature in the CDW state may suggest intrinsic time-reversal symmetry breaking. However, as discussed below, the complex longitudinal and Hall magnetoconductivities can be largely reproduced by multiband analysis.

\begin{figure*}
\centering
\includegraphics[scale=0.6]{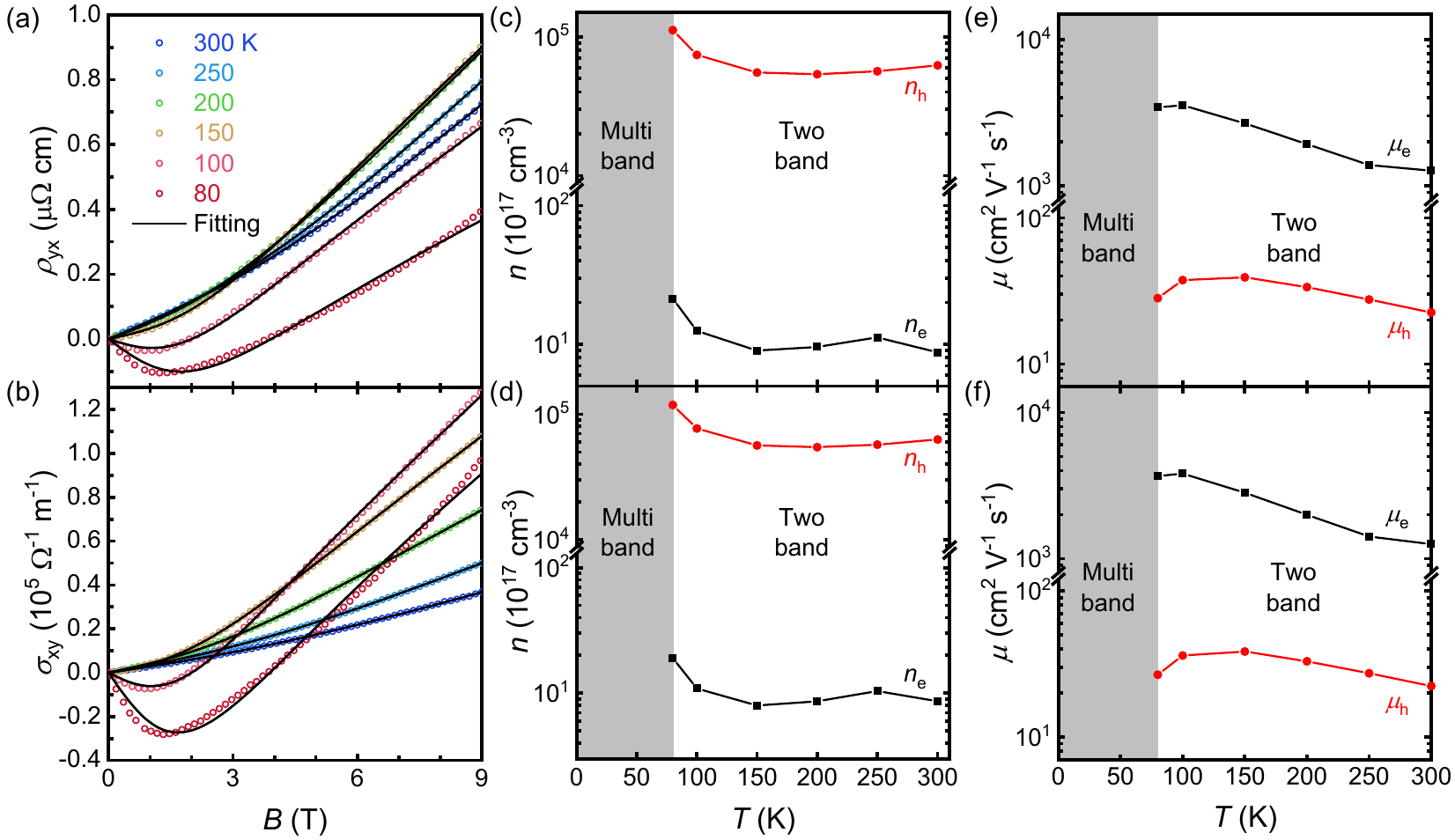}
\caption{(a) and (b) Simple two-band analysis of the high-temperature Hall resistivity and Hall conductivity of \ce{LuNb6Sn6}. The open symbols represent experimental data and the solid lines are fitted curves. (c), (d) and (e), (f) Carrier density and mobility obtained from the simple two-band model using Hall resistivity and Hall conductivity, respectively.}
\label{fig3}
\end{figure*}

To examine the role of multiband transport, we first use a simple two-band model to describe the less complex high-temperature Hall resistivity and Hall conductivity, as shown in Fig.~\ref{fig3}:
\begin{equation}
   \rho_{yx}(B)=\frac{B}{e}\frac{(n_{h}\mu_{h}^{2}-n_{e}\mu_{e}^{2})+\mu_{h}^{2}\mu_{e}^{2}B^{2}(n_{h}-n_{e})}{(n_{h}\mu_{h}+n_{e}\mu_{e})^{2}+\mu_{h}^{2}\mu_{e}^{2}B^{2}(n_{h}-n_{e})^{2}},
   \label{eq1}
\end{equation}
\begin{equation}
    \sigma_{xy}(B)=eB[\frac{n_h\mu_h^2}{1+\mu_h^2B^2}-\frac{n_e\mu_e^2}{1+\mu_e^2B^2}],
    \label{eq2}
\end{equation}
where $n_{e(h)}$ and $\mu_{e(h)}$ represent the carrier density and mobility of the corresponding electronlike (holelike) carriers, respectively. During the fitting, the zero-field longitudinal conductivity, $\sigma_{xx}$($B$ = 0) $=n_h e\mu_h+n_e e\mu_e$, was used as a constraint. As displayed in Figs.~\ref{fig3}(a) and \ref{fig3}(b), the simple two-band model reproduces the measured $\rho_{yx}(B)$ and $\sigma_{xy}(B)$ reasonably well, particularly above 100 K. At 80 K, slight deviations between the two-band fits and the experimental data are observed. Below 80 K, however, the simple two-band model becomes inadequate for describing the complex field dependence of the Hall resistivity and Hall conductivity. From the high-temperature two-band analysis, we extract the carrier density and mobility of the electronlike and holelike carriers, as shown in Figs.~\ref{fig3}(c)--\ref{fig3}(f). The parameters estimated from the Hall conductivity and Hall resistivity are in good agreement. Above 80 K, the Hall resistivity and Hall conductivity are dominated by high-density ($n_h\sim 10^{21}$ cm$^{-3}$) and low-mobility ($\mu_h\sim 10^{1}$ cm$^{2}$V$^{-1}$s$^{-1}$) holelike carriers, leading to the overall positive slope of $\rho_{yx}(B)$ and $\sigma_{xy}(B)$ at high magnetic fields. Meanwhile, the presence of low-density ($n_e\sim 10^{18}$ cm$^{-3}$) and high-mobility ($\mu_e\sim 10^{3}$ cm$^{2}$V$^{-1}$s$^{-1}$) electronlike carriers gives rise to the low-field sublinear behavior and even a negative initial slope in $\rho_{yx}(B)$ and $\sigma_{xy}(B)$.

\begin{figure*}
\centering
\includegraphics[scale=0.52]{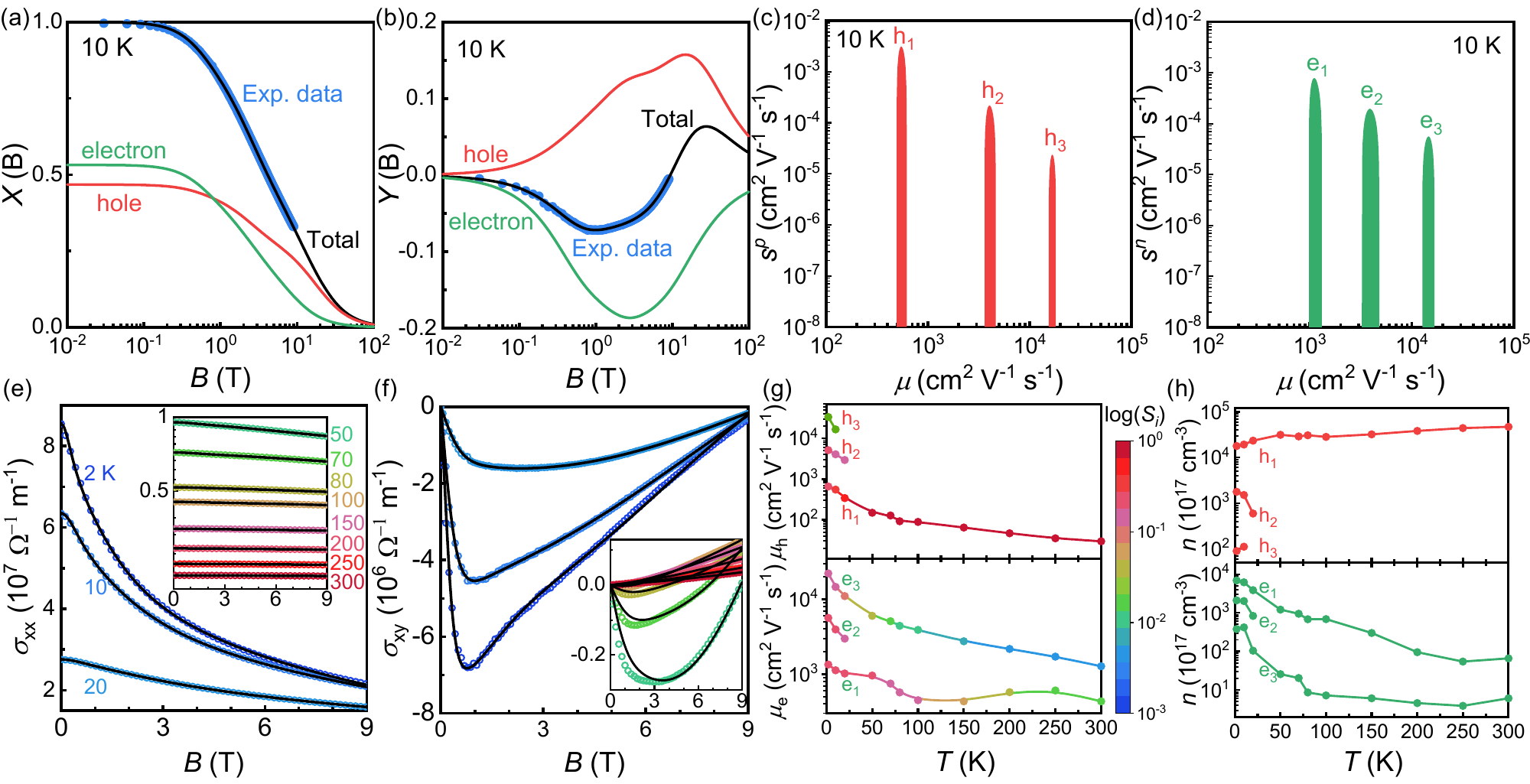}
\caption{(a) and (b) Normalized longitudinal magnetoconductivity $X(B)$ and Hall conductivity $Y(B)$ of \ce{LuNb6Sn6} at 10 K, respectively. The light blue circles represent the experimental data, whereas the black lines correspond to the Lorentzian fitting results. The red and green lines represent the contributions from holelike and electronlike carriers, respectively. (c) and (d) Mobility spectra of hole and electron carriers in \ce{LuNb6Sn6} at 10 K, respectively. (e) and (f) Comparisons between experimental data (open circles) and fitting curves (black lines) for $\sigma_{\mathrm{xx}}(B)$ and $\sigma_{\mathrm{xy}}(B)$, respectively. (g) and (h) Temperature dependence of the mobility, carrier density of each carrier channel. The contribution of each carrier channel to the zero-field conductivity is represented by the logarithmic color scale $\mathrm{log}S_i$.}
\label{fig4}
\end{figure*}

To obtain more quantitative information on the multiband transport, especially in the low-temperature regime where the field dependence of the Hall response becomes highly nonlinear, we employ the mobility-spectrum ($\mu$-spectrum) method. This method does not require \textit{a priori} assumptions on the number or types of charge carriers \cite{Beck_QMSA,Antosz_QMSA,Vurgaftman_QMSA,Liu_MS}. Recently, Liu \textit{et al.} applied this approach to \ce{CsV3Sb5} and showed that its complex magnetotransport can be largely understood within a multiband picture, in which tiny Fermi pockets hosting high-mobility carriers play an important role in producing the AHE-like Hall response \cite{Liu_MS}. Following this approach, we take the data measured at 10~K as a representative example to illustrate the mobility-spectrum analysis.

We first define the normalized longitudinal magnetoconductivity and Hall conductivity as
\begin{equation}
X(B)=\frac{\sigma_{xx}(B)}{\sigma_{xx}(0)},
\end{equation}
and
\begin{equation}
Y(B)=\frac{\sigma_{xy}(B)}{\sigma_{xx}(0)},
\end{equation}
respectively, as shown in Figs.~\ref{fig4}(a) and \ref{fig4}(b). For a multiband system, both quantities contain contributions from holelike and electronlike carriers,
\begin{equation}
X(B)=X^{p}(B)+X^{n}(B),
\end{equation}
and
\begin{equation}
Y(B)=Y^{p}(B)+Y^{n}(B),
\end{equation}
where the superscripts $p$ and $n$ denote hole and electron contributions, respectively.

To separate the two types of carriers, we follow the Kramers--Kronig (KK) procedure introduced in the mobility-spectrum analysis \cite{Liu_MS}. The KK transforms of $X(B)$ and $Y(B)$ are written as
\begin{equation}
X'(B)=\frac{1}{\pi}\mathcal{P}
\int_{-\infty}^{\infty}
\frac{X(B')}{B-B'}\,dB'
=Y^{p}(B)-Y^{n}(B),
\end{equation}
and
\begin{equation}
Y'(B)=\frac{1}{\pi}\mathcal{P}
\int_{-\infty}^{\infty}
\frac{Y(B')}{B-B'}\,dB'
=-X^{p}(B)+X^{n}(B),
\end{equation}
where $\mathcal{P}$ denotes the Cauchy principal value. The hole and electron components are then obtained from
\begin{align}
X^{p}(B)&=\frac{1}{2}\left[X(B)-Y'(B)\right],\\
X^{n}(B)&=\frac{1}{2}\left[X(B)+Y'(B)\right],\\
Y^{p}(B)&=\frac{1}{2}\left[Y(B)+X'(B)\right],\\
Y^{n}(B)&=\frac{1}{2}\left[Y(B)-X'(B)\right].
\end{align}
Since the KK transformation requires knowledge of $X(B)$ and $Y(B)$ over the entire magnetic-field range, the experimental data are first fitted by a sum of Lorentzian functions,

\begin{equation}
X(B)=\sum_{j=1}^{N}
\frac{a_j}{1+\mu_j^2B^2},
\end{equation}

\begin{equation}
Y(B)=\sum_{j=1}^{N}
\frac{b_jB}{1+\mu_j^2B^2}.
\end{equation}
Here, $a_j$, $b_j$, and $\mu_j$ are fitting parameters, and $N$ represents the number of Lorentzian components. In our analysis, $N=5$ provides a sufficiently accurate description of both $X(B)$ and $Y(B)$ over the whole temperature range, while further increasing $N$ does not lead to a significant improvement. We emphasize that $N$ is only a numerical parameter for high-field extrapolation and does not represent the actual number of charge carriers. The number of Lorentzian components used for extrapolation is therefore independent of the number of carrier channels resolved in the mobility spectrum. The fitted $X(B)$ and $Y(B)$ curves are shown together with the experimental data in Figs.~\ref{fig4}(a) and \ref{fig4}(b). 

The obtained Lorentzian functions are subsequently used to perform the KK transformation. Since the practical integration range is finite, the integration is performed within $-10^{6}~\mathrm{T}<B'<10^{6}~\mathrm{T}$, where both $X(B')$ and $Y(B')$ approach zero at the boundaries. This large integration window minimizes numerical artifacts associated with finite-field truncation.

For each carrier type, the separated magnetoconductivities can be expressed in terms of a continuous mobility spectrum,
\begin{equation}
X^{\alpha}(B)
=
\int_{0}^{\infty}
\frac{s^{\alpha}(\mu)}
     {1+\mu^{2}B^{2}}\,d\mu ,
\label{eq:MS_X}
\end{equation}
and
\begin{equation}
Y^{\alpha}(B)
=
\eta_{\alpha}
\int_{0}^{\infty}
\frac{s^{\alpha}(\mu)\mu B}
     {1+\mu^{2}B^{2}}\,d\mu ,
\label{eq:MS_Y}
\end{equation}
where $\alpha=p,n$, $\eta_p=+1$ and $\eta_n=-1$, and $s^{\alpha}(\mu)$ is the normalized conductivity density spectrum. The mobility axis is discretized on an equally spaced logarithmic grid, $m=\ln\mu$, using 1000 grid points. The resulting spectra $s^{p}(\mu)$ and $s^{n}(\mu)$ are obtained by simultaneously fitting $X^{p,n}(B)$ and $Y^{p,n}(B)$.

Figures~\ref{fig4}(c) and \ref{fig4}(d) show the mobility spectra obtained at 10~K. Three characteristic peaks are resolved for both holelike and electronlike carriers, indicating three hole channels ($h_1$, $h_2$, and $h_3$) and three electron channels ($e_1$, $e_2$, and $e_3$). The mobility $\mu_i$ and the contribution $S_i$ of each carrier channel are extracted by integrating the corresponding peak in the mobility spectrum. Since the mobility spectrum is obtained on a discrete mobility grid, the integration is performed numerically as

\begin{equation}
S_i=
\sum_{\mu_i-\epsilon}^{\mu_i+\epsilon}
s(\mu_j)\Delta\mu ,
\end{equation}
where $\Delta\mu$ is the mobility interval and the summation is performed over the region of the $i$-th peak. The averaged mobility of each carrier channel is calculated as
\begin{equation}
\mu_i=
\frac{
\sum_{\mu_i-\epsilon}^{\mu_i+\epsilon}
\mu_j s(\mu_j)\Delta\mu
}
{
\sum_{\mu_i-\epsilon}^{\mu_i+\epsilon}
s(\mu_j)\Delta\mu
}.
\end{equation}
Here, $S_i=\sigma_i(0)/\sigma_{xx}(0)$ represents the fractional contribution of the corresponding carrier channel to the zero-field conductivity, while $\mu_i$ gives the conductivity-weighted average mobility of this channel.

The parameters extracted from the mobility spectra are subsequently used as initial values to simultaneously refine $\sigma_{xx}(B)$ and $\sigma_{xy}(B)$ within a multiband Drude model,
\begin{equation}
\sigma_{xx}(B)
=
\sum_{i\in h}
\frac{n_i e\mu_i}
     {1+\mu_i^2B^2}
+
\sum_{i\in e}
\frac{n_i e\mu_i}
     {1+\mu_i^2B^2},
\label{eq:sigmaxx_multiband}
\end{equation}
and
\begin{equation}
\sigma_{xy}(B)
=
\sum_{i\in h}
\frac{n_i e\mu_i^2 B}
     {1+\mu_i^2B^2}
-
\sum_{i\in e}
\frac{n_i e\mu_i^2 B}
     {1+\mu_i^2B^2}.
\label{eq:sigmaxy_multiband}
\end{equation}
As shown in Figs.~\ref{fig4}(e) and \ref{fig4}(f), the calculated $\sigma_{xx}(B)$ and $\sigma_{xy}(B)$ reproduce the experimental data well over a wide temperature range. This demonstrates that the pronounced complex magnetotransport originates predominantly from the coexistence of several electron and hole channels with strongly different mobilities. Noticeable deviations between the fitted and measured data remain in the intermediate-temperature regime between about 50 and 100 K [inset of Fig.~\ref{fig4}(f)], which will be discussed below.

\begin{figure*}
\centering
\includegraphics[scale=0.6]{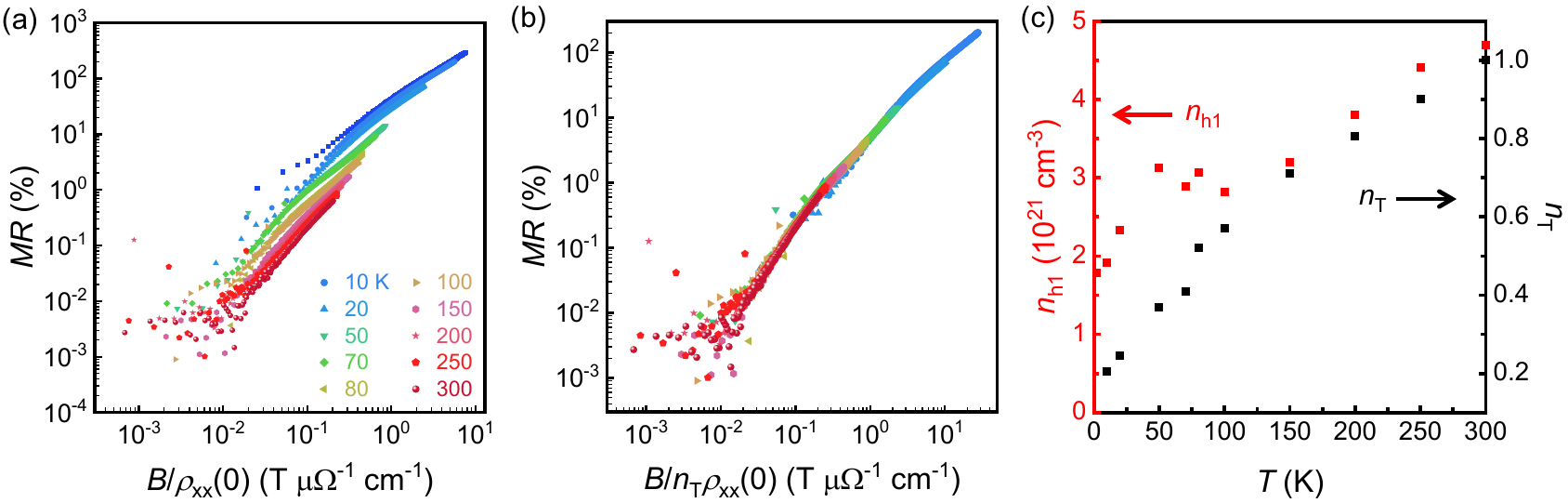}
\caption{(a) Kohler's rule scaling of MR versus $B/\rho_{xx}(0)$ for \ce{LuNb6Sn6}. (b) Extended Kohler's rule scaling with MR plotted against $B/n_\mathrm{T}\rho_{xx}(0)$. (c) Comparison between the temperature-dependent $n_{\mathrm{T}}$ and the carrier density of the dominant holelike channel $n_{\mathrm{h1}}$ obtained from the mobility-spectrum analysis.}
\label{fig5}
\end{figure*}

The carrier density associated with each mobility channel can further be estimated from
\begin{equation}
\sigma_{xx}(0)S_i=n_i e\mu_i,
\end{equation}
yielding
\begin{equation}
n_i=
\frac{\sigma_{xx}(0)S_i}
     {e\mu_i}.
\label{eq:carrier_density}
\end{equation}
The temperature evolution of $\mu_i$, $S_i$, and $n_i$ is summarized in Figs.~\ref{fig4}(g)--\ref{fig4}(i). Above 150 K, the holelike $h_1$ channel dominates the conductivity, with $S_{h1}$ exceeding 90\%. Its high carrier density $n_{h1}\sim10^{21}~\mathrm{cm}^{-3}$ and low mobility $\mu_{h1}\sim10^1$~cm$^{2}$V$^{-1}$s$^{-1}$ are also consistent with the parameters obtained from the simple two-band analysis. Two electronlike channels, $e_1$ and $e_3$, with lower carrier densities but higher mobilities, contribute weakly in the high-temperature regime. Upon cooling below 100 K, the contributions of the $e_1$ and $e_3$ channels increase progressively, with $S_{e1}+S_{e3}$ reaching about 20\% at 50 K. Below 20 K, the mobilities of $e_1$ and $e_3$ reach values on the order of $10^{4}$~cm$^{2}$V$^{-1}$s$^{-1}$, producing steep variations of $\sigma_{xy}(B)$ at small magnetic fields [see Fig.~\ref{fig4}(f)]. Additional channels, $e_2$, $h_2$, and $h_3$, also appear at low temperatures. At 2 K, the total contribution from electronlike channels outweighs that from holelike channels, with $\sum S_{ei}=0.55$ and $\sum S_{hi}=0.45$, respectively. The competition among multiple electronlike and holelike channels with strong mobility contrast can naturally give rise to complex field-dependent magnetotransport.

Additionally, Kohler's rule for magnetoresistance (MR) is violated in \ce{LuNb6Sn6}. Kohler's rule states that the MR of a metal should follow a scaling form, MR$=f[B/\rho_{xx}(0)]$, where MR is a function of $B/\rho_{xx}(0)$ \cite{Kohler}. However, as shown in Fig.~\ref{fig5}(a), the MR curves measured at different temperatures do not collapse onto a single curve when plotted against $B/\rho_{xx}(0)$, indicating a clear violation of Kohler's rule. Such violations have been observed in various materials, ranging from strongly correlated systems to topological semimetals \cite{Harris_vio, Anderson_vio, Xiang_vio, R_vio, Wu_vio, Xu_vio,DeStefano_Sc,Shu_2025}. In many cases, the scaling can be improved using an extended form of Kohler's rule, MR$=f[B/n_\mathrm{T}\rho_{xx}(0)]$, where $n_\mathrm{T}$ represents the temperature-dependent relative variation in carrier density \cite{Xu_vio}. In Fig.~\ref{fig5}(b), we apply this extended Kohler scaling to the MR data of \ce{LuNb6Sn6}. Here, $n_\mathrm{T}$ is set to 1 at 300 K, and its values at other temperatures are adjusted so that the MR curves follow the trend of the 300 K curve. The extracted temperature-dependent $n_\mathrm{T}$ is shown in Fig.~\ref{fig5}(c), together with the temperature-dependent carrier density $n_{h1}$ of the dominant holelike channel. Above 100 K, the trends of $n_\mathrm{T}$ and $n_{h1}$ are close to each other. Deviations between $n_\mathrm{T}$ and $n_{h1}$ at lower temperatures may be associated with CDW-induced band-structure reconstruction and the enhanced contributions of other carriers.

Similar temperature-dependent $n_\mathrm{T}$ and dominant carrier density have been found in \ce{ScV6Sn6}, likely caused by the formation of a pseudogap \cite{DeStefano_Sc}. Indeed, CDW-induced gap opening has been revealed by ARPES on the Sn $5p$ bands of \ce{ScV6Sn6} around $\overline{\Gamma}$ \cite{Lee_ScVSn,Yang_ScVSn,Tuniz_ScVSn}. However, in \ce{LuNb6Sn6}, no clear signature of gap opening has been identified by ARPES experiments \cite{Rui_LuNbSn}. The temperature response of $n_\mathrm{T}$ and the carrier density can arise either from gap opening at the Fermi level during a phase transition or from thermal excitation in semiconductors and semimetals \cite{Xu_vio}. Since no clear gap opening is observed in \ce{LuNb6Sn6}, the temperature-dependent variation in carrier density is more likely related to thermal excitation. Similar thermal excitation-induced temperature-dependent carrier densities have also been found in various multiband metals showing extended Kohler scaling of MR, such as TaP \cite{Xu_vio}, \ce{PdSn2} \cite{PbSn2}, NiSe \cite{NiSe}, \ce{TaCo2Te2} \cite{TaCoTe2}, and the kagome metal \ce{EuTi3Bi4} \cite{Shu_2025}. 

Finally, it is worth noting that systematic deviations between the multiband analysis and the experimental data remain in the intermediate temperature range of approximately 50--100~K. Increasing the number of field-independent Drude channels does not remove these deviations. Since this temperature window encompasses $T_{\mathrm{CDW}}\approx80$~K, the discrepancy likely reflects additional CDW-related modifications of quasiparticle scattering, such as field-dependent or strongly momentum-dependent scattering, which are not captured by a conventional mobility-spectrum analysis with field-independent mobilities. Elastic x-ray scattering studies have revealed two competing orders in the vicinity of $T_{\mathrm{CDW}}$ \cite{Yang_LuNbSn}. In particular, a phonon instability at the wave vector $\mathbf{Q}_\mathrm{H}=(1/1, 1/3, 1/2)$ develops gradually below about 160 K, whereas the CDW instability appears abruptly below $T_{\mathrm{CDW}}$ at a distinct wave vector $\mathbf{Q}_\mathrm{CDW}=(1/3, 1/3, 1/3)$ \cite{Yang_LuNbSn}. The development of competing orders and related charge fluctuations, together with CDW-induced band-structure reconstruction, may strongly reshape quasiparticle scattering and make the simple assumption of field-independent mobilities inapplicable in this critical temperature range. On the other hand, the possibility of time-reversal-symmetry breaking effects cannot be fully excluded in view of the discrepancy between the multiband analysis and the observed complex magnetotransport near the CDW transition. In the kagome metals \textit{A}\ce{V3Sb5} and \ce{ScV6Sn6}, time-reversal symmetry breaking effects have been identified by various experimental techniques \cite{Jiang2021a, Wang2021CDW, Shumiya2021, Yang2020, Yu2021, Zheng2021_gating, Chendong2021, Zhou2021, Mielke2021, Yu2021b, Gan2021, Mi_2022, Liang2021CDW, Zhao2021a, Chen2021rotonpair, Xu2021, Li2021a, Hu2022co,Guguchia2023_ScVsn}. It would therefore be interesting to explore the possibility of similar effects in \ce{LuNb6Sn6} using other probes, such as muon spin relaxation, Kerr effect, and scanning tunneling microscopy.

\section{Conclusions}

In summary, we have studied the magnetotransport properties of the kagome metal \ce{LuNb6Sn6}, which undergoes a CDW transition at 79 K. Both the longitudinal and Hall magnetoconductivities exhibit complex field-dependent behavior. At low temperatures, the low-field longitudinal magnetoconductivity shows an approximately linear dependence on magnetic field, while the Hall magnetoconductivity undergoes a sign change at a temperature-dependent magnetic field, similar to the behavior observed in \textit{A}\ce{V3Sb5} and \ce{ScV6Sn6}. Mobility-spectrum analysis reproduces these complex behaviors well at both low and high temperatures, highlighting the important role of multiband transport. In the intermediate-temperature regime of about 50--100 K near the CDW transition, however, the complex magnetotransport cannot be fully described by simple multiband transport with field- and momentum-independent mobilities. These discrepancies may arise from momentum-dependent scattering induced by competing charge orders and/or possible time-reversal symmetry breaking effects.

\section{Acknowledgments}

We thank Yan Liu and Guiwen Wang at the Analytical and Testing Center of Chongqing University for their technical support.  This work was supported by the National Natural Science Foundation of China (Grant No. 12474141), the Natural Science Foundation of Chongqing, China (Grant No. CSTB2025NSCQ-GPX0729), the Fundamental Research Funds for the Central Universities, China (Grant No. 2025CDJ-IAISYB-034), the Venture and Innovation Support Program for Chongqing Overseas Returnees (Grant No. cx2024007), and the Chinesisch-Deutsches Zentrum für Wissenschaftsförderung through the Chinesisch-Deutsche Mobilitätsprogramm (Grant No. M-0496). Y.C. acknowledges support from the National Natural Science Foundation of China (Grant Nos. 12374081 and 12227806), the Open Research Fund of the Pulsed High Magnetic Field Facility (Grant No. WHMFC2024007), and Huazhong University of Science and Technology.

\end{document}